\documentclass[11pt,oneside,listsleft]{scrartcl}
\usepackage{natbib}
\usepackage{graphicx}
\usepackage{ulem}
\usepackage{amssymb,amsmath}
\usepackage{booktabs,units}
\usepackage[latin1]{inputenc}
\usepackage{scrpage2}
\usepackage{soul}
\usepackage{setspace}
\usepackage{longtable}
\usepackage{caption}

\begin{document}
\thispagestyle{empty}

\begin{center}
{\LARGE\sffamily\bfseries Reconstructing the Traffic State by\\[1ex]Fusion of Heterogeneous Data}\\[4ex]

{\large\sffamily\bfseries{Martin~Treiber$^\text{a,1}$,  Arne~Kesting$^\text{a,2}$}, and R.~Eddie~Wilson$^\text{b,3}$ }
\end{center}

\begin{itemize}
\item[$^\text{a}$]Institute for Transport \& Economics, Technische
Universit\"at Dresden, Falkenbrunnen, W\"urzburger Str. 35, 01187 Dresden (Germany)
\item[$^\text{b}$]Department of Engineering Mathematics, University of
  Bristol, Queen's Building, Bristol BS8 1TR (UK)
\footnotetext[1]{\sf E-mail: {\tt treiber@vwi.tu-dresden.de}, URL: {\tt www.mtreiber.de}}
\footnotetext[2]{\sf E-mail: {\tt kesting@vwi.tu-dresden.de}, URL: {\tt www.akesting.de}}
\footnotetext[3]{\sf E-mail: {\tt Re.Wilson@bristol.ac.uk}, URL: {\tt http://www.enm.bris.ac.uk/staff/rew}}
\end{itemize}

\section*{Abstract:}
We present an advanced interpolation method for estimating smooth
spatiotemporal profiles for local highway traffic variables such as
flow, speed and density. The method is based on stationary detector
data as typically collected by traffic control centres, and may be
augmented by floating car data or other traffic information. The
resulting profiles display transitions between free and congested
traffic in great detail, as well as fine structures such as
stop-and-go waves.  We establish the accuracy and robustness of the
method and demonstrate three potential applications: 1.\ compensation
for gaps in data caused by detector failure; 2.\ separation of noise
from dynamic traffic information; and 3.\ the fusion of floating car
data with stationary detector data.

\section{\label{sec:intro}INTRODUCTION}

A detailed picture of speed and flow is essential for understanding
flow breakdown on highways and the dynamics of congestion. In
particular, highway traffic may not be understood by time series data
alone but rather we must consider its structure in space and time
jointly \citep{Opus,Bertini-DataFusion05,Bertini-BottleneckAnalysis05}.  
Let $t$ and $x$ denote respectively time and distance driven
down the highway. We may thus introduce {\it spatiotemporal profiles}
for the macroscopic variables velocity $V(x,t)$, flow $q(x,t)$ and
density $\rho(x,t)$. Where the quality of data permits, these
quantities may be displayed as colour charts or landscapes which
exhibit rich structure such wave propagation, phase transitions
etc. To continue the analogy with particle physics, we may speak of
the {\it traffic state}, by which we mean the classification of the
traffic at any one time according to its spatiotemporal structure,
e.g., as {\it free flow}, {\it synchronized flow}, {\it stop-and-go
  waves} \citep{Kerner-Rehb96}, or in terms of more detailed
classifications \citep{Phase,Helb-Phases-EPJB-09}.

Unfortunately, highway traffic data comes from many heterogeneous
sources \citep{lintCACAIE2009}.  Most simply we have stationary
detector data (SDD) collected by fixed infrastructure, which typically
consists of inductance loops buried in the surface of the road. In
their usual operation, the loops count vehicles and estimate their
lengths and speeds, which are then sent to regional traffic control
centres in the form of 1-minute aggregate data.  However, modern
communication systems have sufficient bandwidth to carry full
Individual Vehicle Data (IVD) to the control centre, and this data may
lead to advances in incident detection algorithms, for example. More
recent stationary detection systems operate in a similar fashion to
inductance loops, but are based on magnetometers, and radar / laser /
infrared devices installed on bridges.

Other traffic data is not provided at fixed points in space, for
example floating-car data (FCD) from GPS devices
\citep{fastenrath1997floating,Ivan-DataFusion98,Herrera-Reconstruction08},
or floating-phone data (FPD)
\citep{caceres2008review}. Yet further data is produced on an
event-oriented basis such as messages from the police.

Two problems with such highway traffic data sets are:
\begin{enumerate}
\item Sparseness. Each source of data individually may be insufficient
  to determine the traffic state. For example, the distance between
  consecutive stationary detectors may be too great to infer what is
  happening between them. This problem is compounded by detector
  failure.  In the case of FCD, it is a relatively small proportion of
  vehicles that report GPS data.
\item Noise, of several different types, for example: 1.\ measurement
  error committed by detectors; 2.\ sampling errors in aggregate data
  due to small numbers (a problem in low flow conditions); and
  3. heterogeneity of the driver / vehicle population (so that FCD for
  one vehicle may not be at all representative of those around it).
\end{enumerate}
However, both sparseness and noise can be addressed to some degree by
combining data from either similar or heterogeneous sources, and the
focus of this paper is a method which can fuse heterogeneous data
in order to reconstruct smooth spatiotemporal profiles for local
traffic variables such as the speed. 

Our method is based on the {\it adaptive smoothing method} (ASM)
\citep{Treiber-smooth}. Whereas the original ASM was restricted to SDD
only, here we consider the {\it generalized} adaptive smoothing method
(GASM) which is
extended to cope with heterogeneous sources~\citep{Kesting-Traveltime-Fovus08}. In particular, the GASM
 can interpolate locally inconsistent data: for example, an FCD speed
measurement need not equal a 1-minute aggregate SDD with the same
$(x,t)$ coordinates, but the GASM can nevertheless smoothly combine
these disparate measurements.

Both the ASM and GASM interpolate between data points (thereby
tackling the problem of incomplete data coverage) and eliminate
high-frequency noise while preserving most of the relevant dynamic
information. However, their chief novelty is that they surpass
non-specialized smoothing methods by incorporating some traffic
physics, in the form of well-understood wave propagation
characteristics \citep{Helb-Phases-EPJB-09}. 
Specifically, it is known that in congestion, small
perturbations to the traffic propagate upstream at about
\unit[15]{km/h}, whereas in free-flow they propagate downstream at the
speed of the vehicles \citep{Kerner-Rehb96,CasBer-99,Martin-empStates}.

The chief contributions of this paper are an investigation of the
robustness and accuracy of the GASM and a demonstration of some
potential applications, including the fusion of FCD and SDD.  The
organization is as follows. In Sec.~\ref{sec:ASM}, we formulate the
GASM and discuss calibration and validation issues based on data from
the German autobahn A9 and the English M42 motorway.  Since the
detector coverage on the M42 is extremely dense (loop detector
spacings of \unit[100]{m}), the validation (Sec.~\ref{sec:validation})
will be based on real (not simulated) traffic dynamics which can be
considered as completely known for our purposes.  In
Sec.~\ref{sec:appl}, we propose possible applications of this method
such as bridging data gaps (Sec.~\ref{sec:dataGaps}), separating noise
from information (Sec.~\ref{sec:noise}), and fusion of floating-car
and stationary detector data (Sec.~\ref{sec:fcd}).  Finally, the
method and the results will be briefly discussed in
Sec.~\ref{sec:disc}.

\section{\label{sec:ASM}GENERALIZED ADAPTIVE SMOOTHING METHOD}

We now present the details of the generalized adaptive smoothing
method (GASM) which performs two-dimensional interpolation to
reconstruct the spatiotemporal traffic state from discrete traffic
data. Speed data $v_i$ measured at known locations $x_i$ and times
$t_i$ are obtained from either SDD or FCD, or on an event-oriented
basis, and are combined to produce a smooth velocity field as a
function of continuous space and time. With minor modifications, flow
and other spatiotemporal variables may also be reconstructed.

The GASM is based on two-dimensional interpolation in space and time
(Sec.~\ref{sec:interpolation}) using smoothing kernels. However, in
contrast to a conventional isotropic filter, the method incorporates
the known characteristic velocities of information propagation in free and
congested traffic (Sec~\ref{sec:anisotropic}), by skewing the
principal axes of the smoothing kernel. The switch between free and
congested traffic is then managed by a nonlinear adaptive speed filter
(Sec.~\ref{sec:vFilter}). We then demonstrate the effectiveness of the
GASM by comparing it with conventional smoothing with an isotropic
kernel (Sec.~\ref{sec:exampleA9}). Finally, in
Sec.~\ref{sec:validation}, we validate the GASM using M42 data where
the inductance loop system is over-specified. The approach is to apply
the GASM to a subset of the inductance loop data and re-construct the
velocity field at the positions of detectors which have not been used 
in the interpolation. The accuracy of the GASM may then be established in 
comparison to the detector data which is regarded as the ground truth.

\subsection{\label{sec:interpolation}Conventional Spatiotemporal Interpolation}
Our inputs are a set of discrete data points $\{x_i,t_i,v_i\}$,
$i=1,\ldots,n$, and the interpolation task is to derive from them a
smooth velocity field $V(x,t)$ in a given spatiotemporal interval. The
broad approach is to employ the convolution
\begin{equation}\label{eq:conv_asm}
V(x,t) = \frac{1}{\mathcal{N}(x,t)} \sum_i \phi_i\left(x-x_i,t-t_i\right)\, 
v_i,
\end{equation}
where the smoothing kernels $\phi_i(x,t)$ are sufficiently localized
functions that decrease with increasing $|x|$ or $|t|$, and we
define the normalization factor by
\begin{equation}
\mathcal{N}(x,t) = \frac{1}{\sum\limits_i \phi_i(x-x_i, t-t_i)}.
\end{equation}
This formulation allows for different types of data point to use different
kernels, but to simplify matters we shall usually assume that the kernels
are identical and take the symmetric exponential form 
\begin{equation}\label{eq:iso_kernel} 
\phi(x, t) = \exp\left[  -\left( \frac{| x|}{\sigma} + \frac{|t|}{\tau} \right) \right],
\end{equation}
although a bivariate Gaussian would also be suitable. The smoothing
kernel acts as a kind of {\it low-pass filter}, and the positive
constants $\sigma$ and $\tau$ define characteristic `widths' for
the spatial and temporal smoothing respectively, so that features with
finer scales tend to be smoothed out.

Suitable values for $\sigma$ and $\tau$ are of the order of half the
typical distance between neighboring data points.  For example, in a
typical situation where an inductance loop system provides one-minute
aggregate data at \unit[2]{km} intervals, we choose
$\tau=\unit[30]{s}$ and $\sigma=\unit[1]{km}$ (see
Sec.~\ref{sec:validation}). However, for a stronger reduction of
noise, larger smoothing widths may be chosen (e.g., $\tau$ up to
\unit[2]{min}).
\subsection{\label{sec:anisotropic}Traffic-Adaptive Smoothing}

A serious challenge in traffic data is that the typical scale of some
traffic patterns, such as the wavelength of stop-and-go waves, is 
(at \unit[1-2]{km}) similar to the spacing of stationary
detectors. Consequently, important dynamical features may be lost
in the interpolation process, and even entirely spurious patterns
may be reconstructed (see Fig.~\ref{fig:example_A9}).

To enhance the resolution of the filter, we may use established facts
concerning the propagation of information in traffic flow.  The
`propagation velocity' is that at which small perturbations to the
traffic flow are propagated, and in traffic theories based on
hyperbolic partial differential equations, it corresponds to the
characteristic wave velocity given by the gradient of the equilibrium
flow-density curve (the so-called `fundamental diagram'). It is
well-known that:
\begin{enumerate}
\item In free traffic, perturbations move downstream (i.e., in the
  direction of traffic flow) \citep{Kerner-Rehb96,Martin-empStates}
  and the characteristic propagation velocity $c_\text{free}$ is similar to the
  local average speed of the vehicles.
\item In congested traffic, however, perturbations travel {\it
  against} the movement of the vehicles (i.e., upstream). Moreover, the
  characteristic propagation velocity $c_\text{cong}$ is usually in
  the vicinity of \unit[-15]{km/h}. This value is well-established as
  the typical velocity of stop-and-go waves
  ~\citep{Kerner-Rehb96,Martin-empStates}, but seems to apply more
  generally to nearly all information propagation in congested traffic.
\end{enumerate}

\begin{figure}[th]
\centering
 \includegraphics[width=8cm]{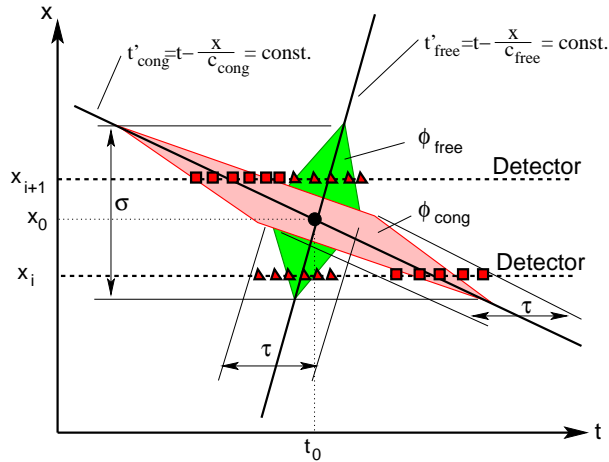} 
 \caption{\label{fig:asm_kernels}Illustration of the smoothing kernels
   for free and congested traffic. The inclination angles capture the
   different characteristic velocities $c_\text{free}$ and
   $c_\text{cong}$. In particular, perturbations propagate upstream
   (against the driving direction) in congested traffic.}
\end{figure}

To apply these facts, we skew the conventional isotropic smoothing
kernel \eqref{eq:iso_kernel} in order to obtain the anisotropic
interpolation formulae
\begin{eqnarray}\label{eq:asm_v_free}
V_\text{free}(x, t) &=& \frac{1}{\mathcal{N}(x,t)}\sum_i \phi\left(x-x_i,
t-t_i-\frac{x-x_i}{c_\text{free}} \right)\, v_i\,,\\
\label{eq:asm_v_cong}
V_\text{cong}(x, t) &=& \frac{1}{\mathcal{N}(x,t)}\sum_i \phi\left(x-x_i,
t-t_i-\frac{x-x_i}{c_\text{cong}} \right)\, v_i\,,
\end{eqnarray}
for free and congested traffic respectively.  In effect, these new
filters correspond to smoothing in preferred directions in the $(x,t)$
plane (see Fig.~\ref{fig:asm_kernels}), based on the propagation
velocities $c_\text{free}$ and $c_\text{cong}$. Note that the
conventional isotropic smoothing corresponds to the limit
$c_\text{free}\ = c_\text{cong} \to \infty$.

In practice, we take $c_\text{cong} \approx -\unit[15]{km/h}$, whereas
we find that $c_\text{free}\simeq +\unit[70]{km/h}$ gives good results
in a highway context. The use of just two characteristic propagation
velocities is consistent with kinematic wave models~\citep{CasWin95,CasBer-99} and
the assumption of a piecewise linear (`triangular') fundamental
diagram~\citep{New93}.

\subsection{\label{sec:vFilter}Nonlinear Adaptive Filter}
We must now construct a single smoothing filter which combines
the formulae for free and congested traffic.
To this end, we define 
\begin{equation}\label{eq:asm_super}
V(x, t) = w (x,t) \, V_\text{cong}(x, t) + \left[ 1- w(x,t) \right] \,V_\text{free}(x, t)\,,
\end{equation}
where the weight factor $w(x,t)
=W\left(V_\text{free}(x,t),V_\text{cong}(x,t\right))$ controls the
superposition of the free and congested velocity fields
\eqref{eq:asm_v_free},\eqref{eq:asm_v_cong}. We require $W \approx 0$ at high
speeds and $W \approx 1$ at low speeds, and thus we use the smooth $s$-shaped function
\begin{equation}\label{eq:asm_w_filter}
 W(V_\text{free}, V_\text{cong}) = \frac{1}{2}
\left[ 1 + \tanh\left( \frac{V_\text{thr} -  \min \left(V_\text{free},V_\text{cong}\right)}{\Delta
 V} \right)\right].
\end{equation}
Here, the `predictor' $\min(V_\text{free},V_\text{cong})$ is defined
such that the patterns of congested traffic are better reproduced by
the resulting non-linear filter than that for free traffic. The
threshold between free and congested traffic is defined by
$V_\text{thr}$ while the transition width is determined by $\Delta
V$. Typical parameter values are given in Table \ref{tab:asm_param}.

\subsection{\label{sec:exampleA9}Sensitivity Analysis}

For illustration, we apply the GASM to a small portion of one-minute
aggregate detector data from the South-bound A9 autobahn near Munich,
Germany \citep{Opus,Treiber-smooth}.  We consider
data from 9 detectors spread over 14km of highway during 4 hours of a
busy morning on which there were pronounced stop-and-go
patterns. Here, as throughout the paper, 1-minute aggregate SDD
  points also incorporate a flow-weighted aggregate of speed
  measurements across the lanes of the highway. In fact, in congested
  traffic, the speed variance between lanes tends to be rather small,
  and aggregation across lanes thus helps reduce sampling noise.

\begin{figure}
\centering
\begin{tabular}{cc}
\includegraphics[width=59mm]{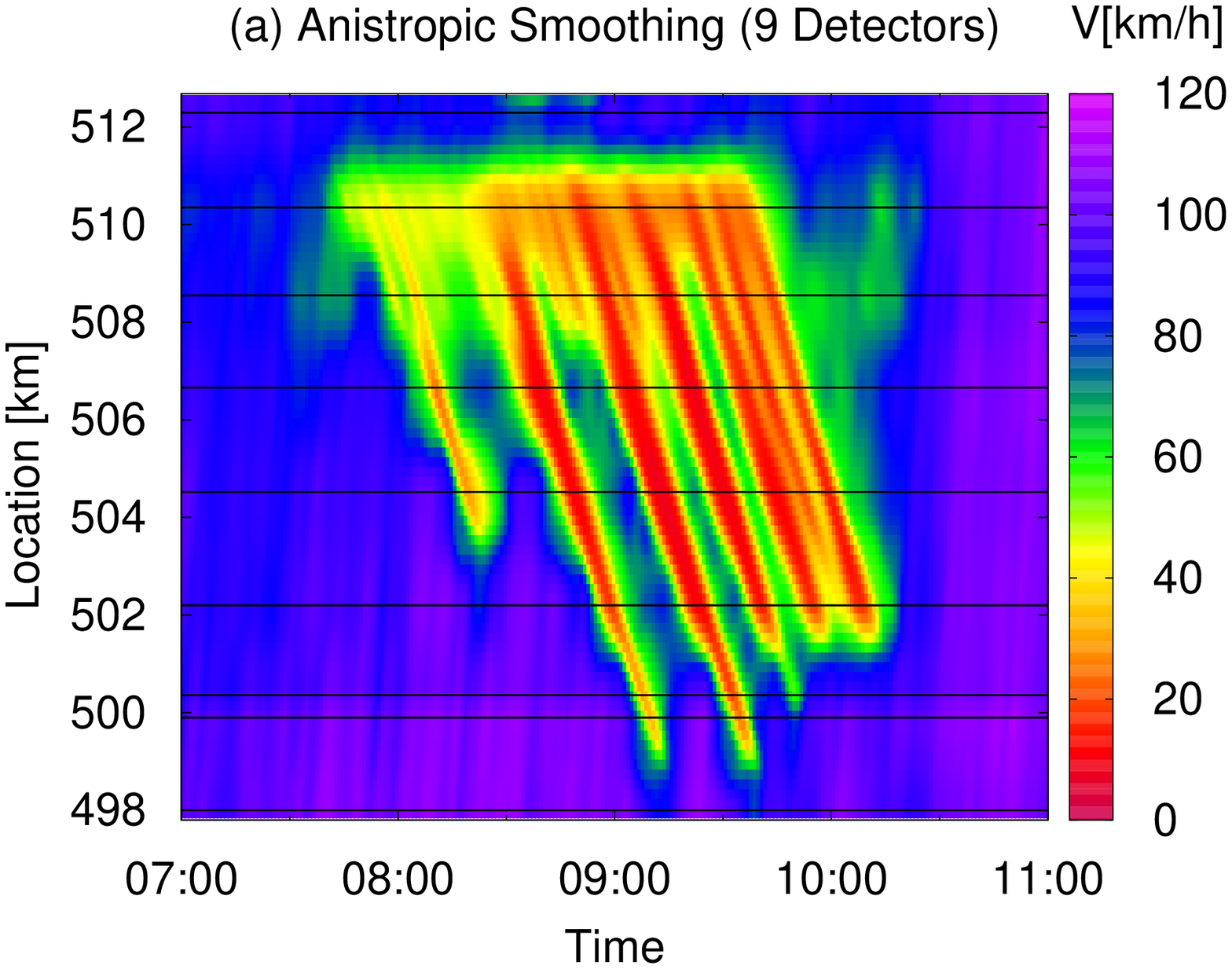}  &
\includegraphics[width=59mm]{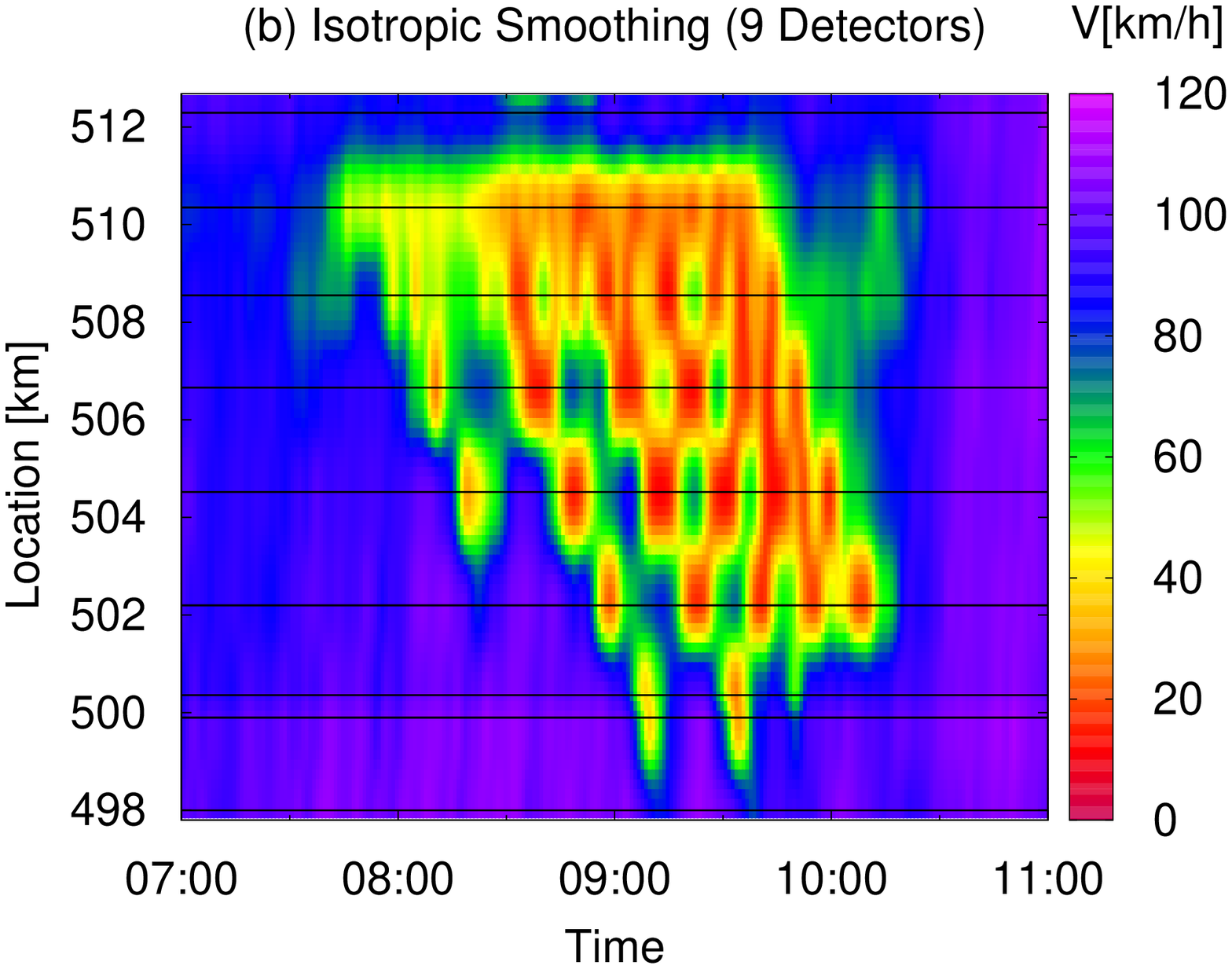}\\ [-5mm]
\includegraphics[width=59mm]{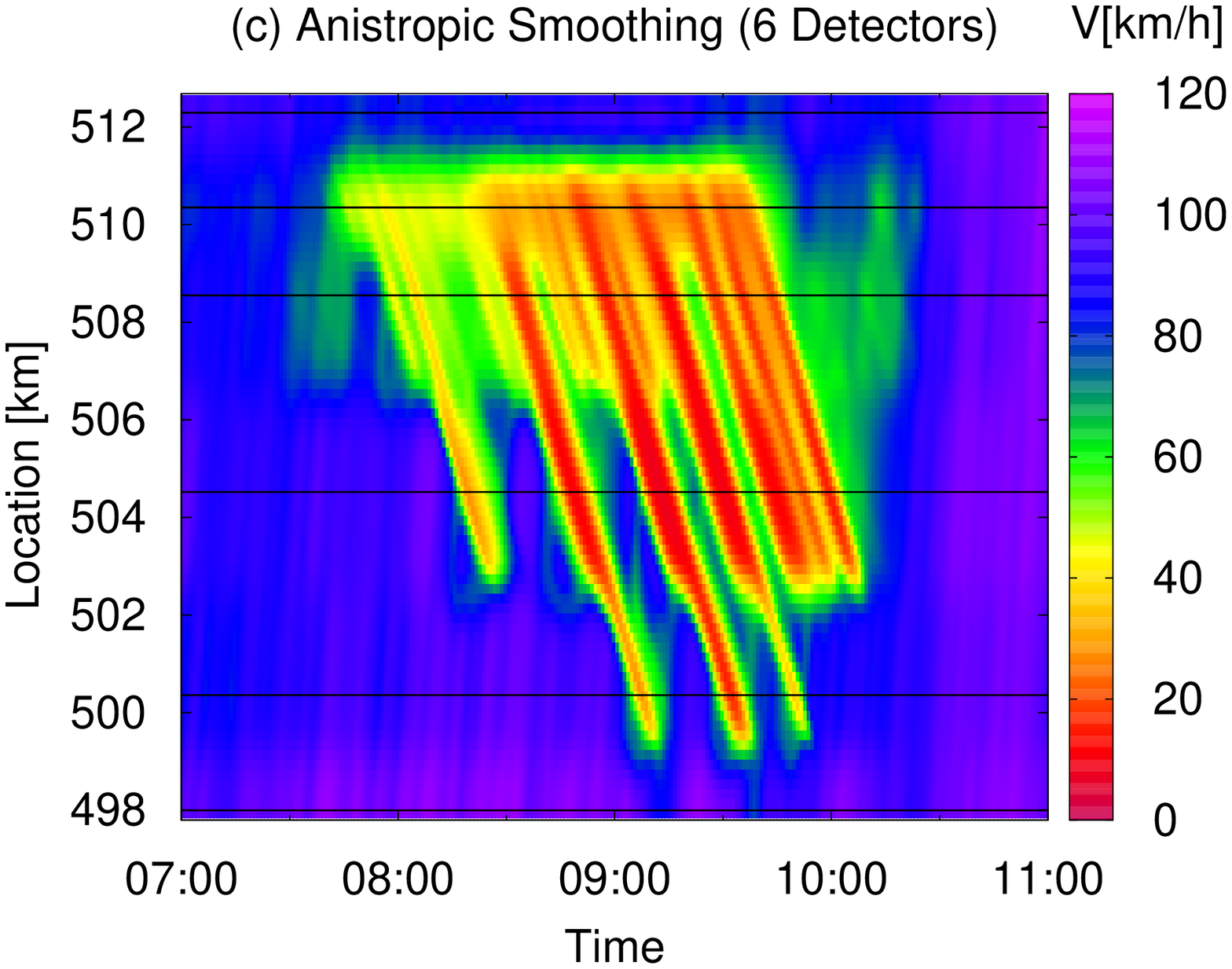} &
\includegraphics[width=59mm]{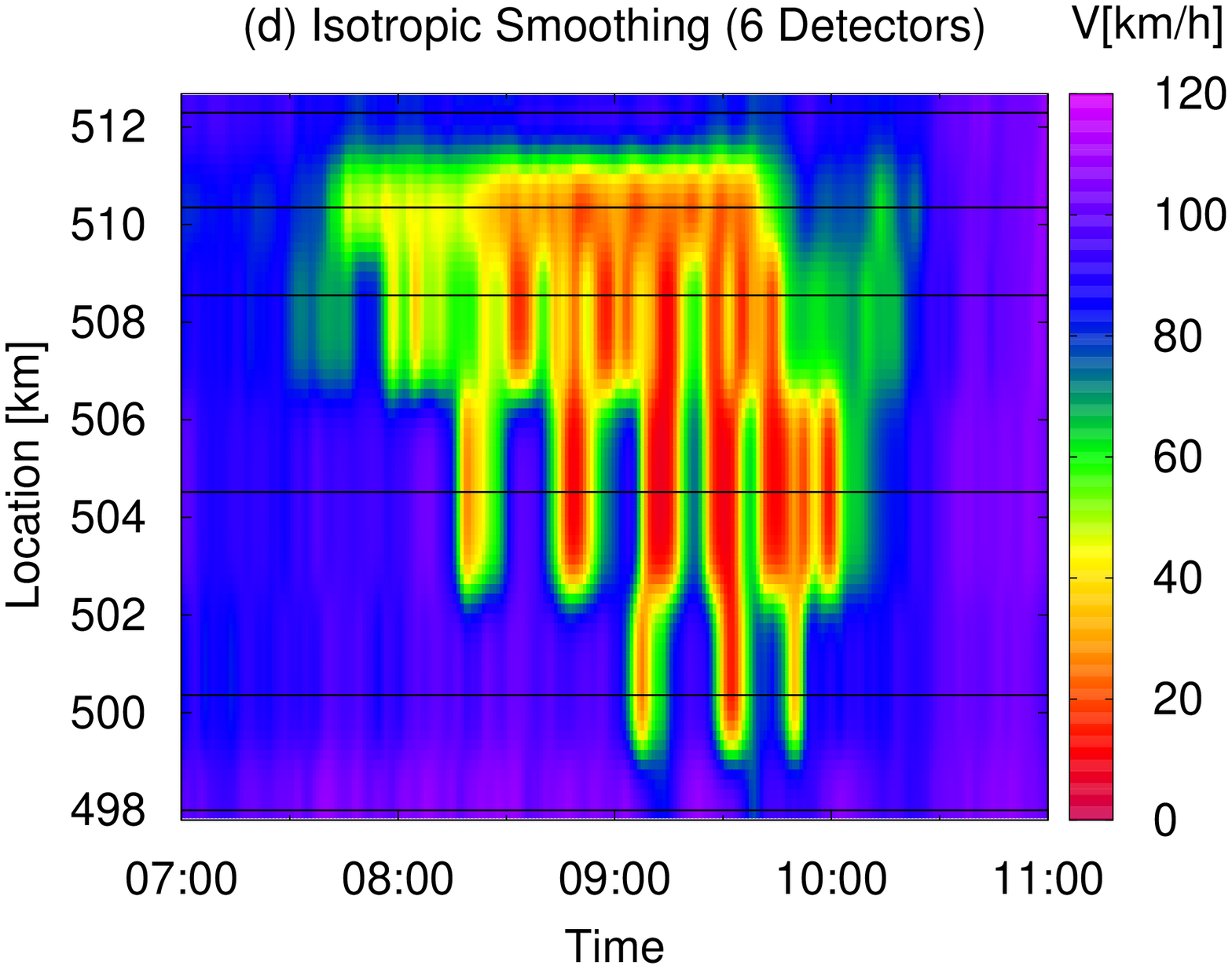}
\end{tabular}
 \caption{\label{fig:example_A9}Generalized Adaptive Smoothing Method
   (a,c) versus conventional isotropic interpolation (b,d) applied to
   loop detector data from the German autobahn A9 (near Munich,
   direction South). In (c,d), only six out of the nine detectors are used
   in reconstruction.  Conventional interpolation may neglect
   important spatiotemporal features or even identify spurious
   structure.}
\end{figure}

See Fig.~\ref{fig:example_A9}, which compares the performance of GASM
with the standard isotropic filter \eqref{eq:conv_asm}. Here GASM uses
the standard parameters from Table~\ref{tab:asm_param}. The positions
of the detectors which are used in reconstruction are indicated by
horizontal lines. Compare Figs.~\ref{fig:example_A9}(a,b): we may
observe that GASM is able to resolve individual stop-and-go waves when
isotropic smoothing is not able to identify the pattern. As a more
difficult challenge, in Figs.~\ref{fig:example_A9}(c,d) the smoothing
algorithms are compared when data from only 6 out of the 9 detectors
is used. In this case isotropic smoothing identifies spurious patterns
although GASM continues to reconstruct stop-and-go waves correctly
even though the spatial resolution of the input data is very poor.

\begin{table}
\centering
{\small
\begin{tabular}{ll}
\toprule Parameter & Value\\ \midrule Smoothing width in space
coordinate $\sigma$ & $\Delta x/2$\\ Smoothing width in time
coordinate $\tau$ & $\Delta t/2$ \\ Propagation velocity of
perturbations in free traffic $c_\text{free}$ &
\unit[70]{km/h}\\ Propagation velocity of perturbations in congested
traffic $c_\text{cong}$ & \unit[-15]{km/h}\\ Crossover from congested
to free traffic $V_\text{thr}$ & \unit[60]{km/h}\\ Transition width
between congested and free traffic $\Delta V$ &
\unit[20]{km/h}\\ \bottomrule
\end{tabular}
}
   \caption{\label{tab:asm_param}Parameters of the adaptive smoothing
     algorithm with typical numerical values used in this paper. The
     spatial and temporal smoothing widths are chosen as half of the average
     inter-detector spacing $\Delta x$ and sampling time $\delta t$,
respectively.} 
\end{table}

\begin{figure}
\centering
\begin{tabular}{cc}
\includegraphics[width=59mm]{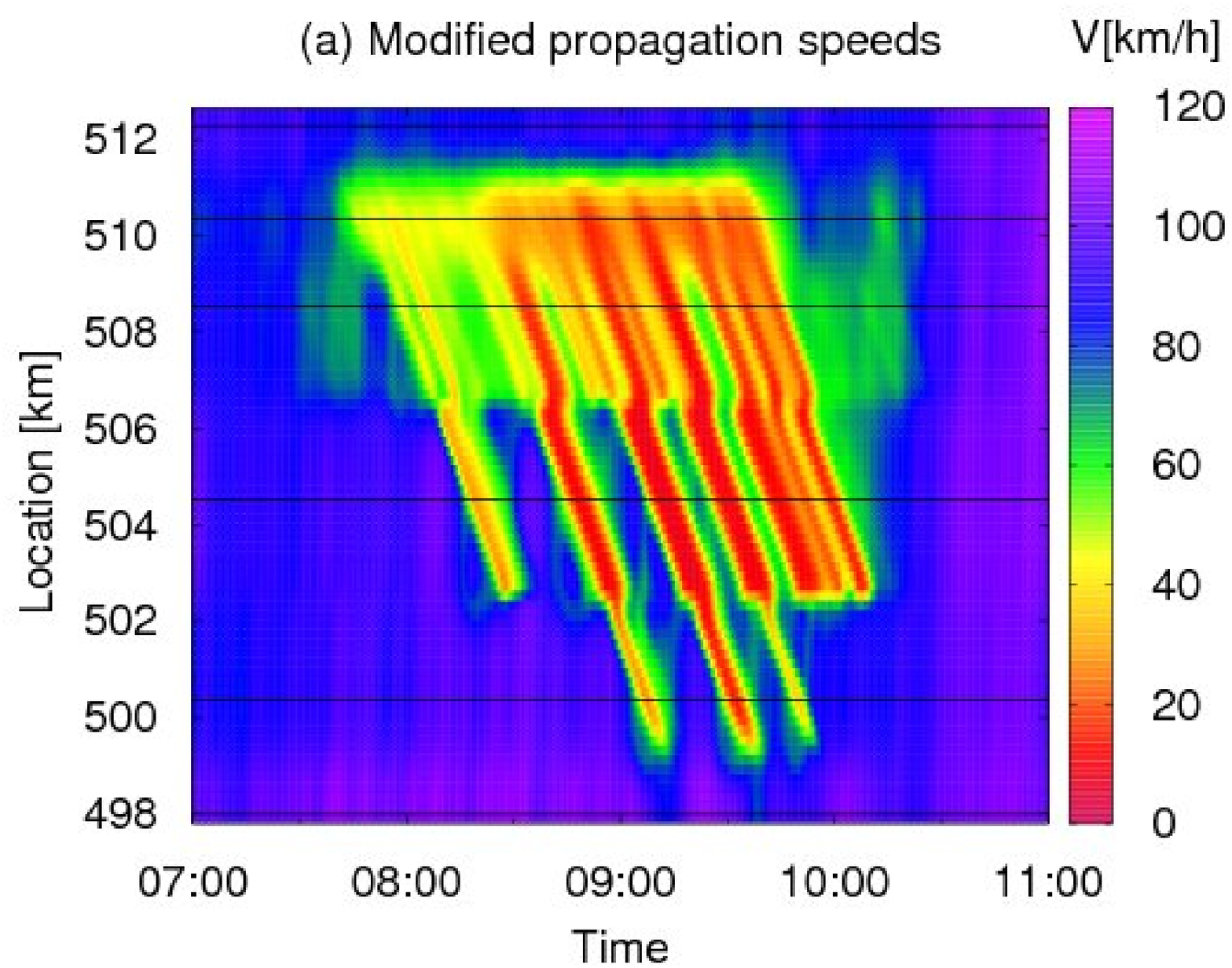}  &
\includegraphics[width=59mm]{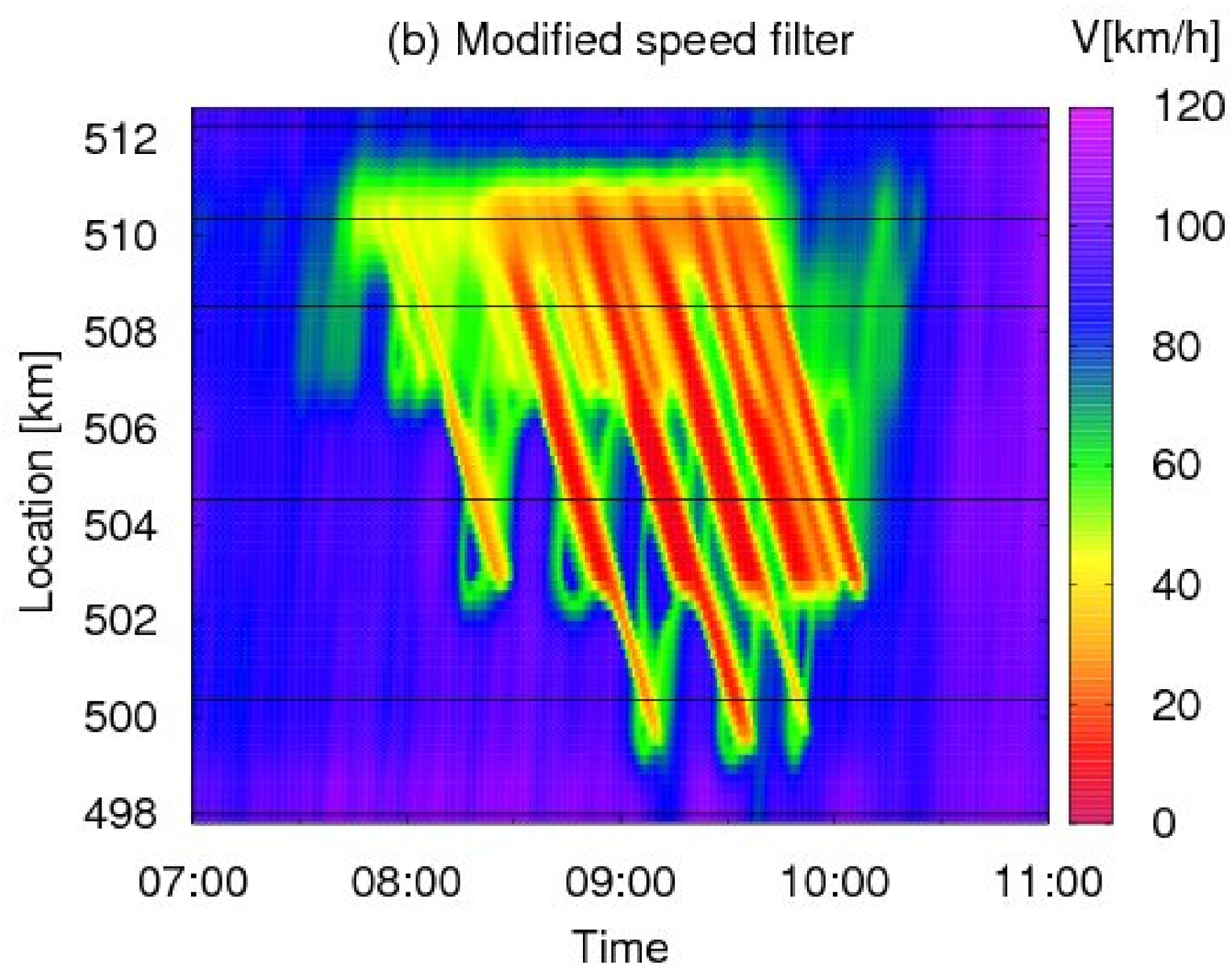}
\end{tabular}
\caption{\label{fig:param_A9}Sensitivity of the generalized adaptive
  smoothing method to variation in its parameters
  (cf. Fig.~\ref{fig:example_A9}, full details in the main text).}
\end{figure}

Our experience is that GASM performs better than isotropic smoothing
in all of the data sets that we have tried. Moreover, we have found
that the parameter choices in Table~\ref{tab:asm_param} are robust and
do not need re-tuning for each new application. As an illustration,
we apply GASM to the same data as for Fig.~\ref{fig:example_A9}, but
with pathological changes to the algorithm parameters. See
Fig.~\ref{fig:param_A9}. 

In Fig.~\ref{fig:param_A9}(a), we change the propagation velocities
$c_\text{free}$ from $\unit[70]{km/h}$ to $\unit[200]{km/h}$, and
$c_\text{cong}$ from $\unit[-15]{km/h}$ to $\unit[-12]{km/h}$. In
contrast, in Fig.~\ref{fig:param_A9}(b), we modify the nonlinear
filter, by reducing the transition width from $\Delta
V=\unit[20]{km/h}$ to $\unit[5]{km/h}$, and the crossover threshold
from $V_\text{thr}=\unit[60]{km/h}$ to $\unit[45]{km/h}$.  To assess
the robustness, these plots should be compared with the corresponding
results for GASM with standard parameters and for conventional isotropic
smoothing, in Figs.~\ref{fig:example_A9}(c,d) respectively.

On visual inspection, the quality of the GASM depends only weakly on
its parameters and in all cases surpasses isotropic smoothing at this
spatial resolution. We have found that the most sensitive parameter is
the propagation velocity $c_\text{cong}$ for congested traffic, since
too low or too high values result in step-like artefacts in the
reconstruction. However, in practice, $c_\text{cong}$ varies very
little from situation to situation~\citep{Kerner-Rehb96,Martin-empStates}.

In conclusion, we have yet to perform a formal optimisation of the
GASM parameters. But our experience is that GASM reconstructs traffic
patterns robustly with the parameter choices of
Table~\ref{tab:asm_param} and stationary detector spacings up to
about~\unit[3]{km}. Near a bottleneck, this spacing should preferably
be reduced so that the stationary downstream jam front can be
accurately positioned.

\subsection{\label{sec:validation}Validation}
For validation of the GASM, we consider a 15-kilometer long section of
the North-bound M42 motorway near Birmingham, England.  As part of the
English Highways Agency's Active Traffic Management system~\citep{EHAATM}
this highway has been equipped with an almost unprecedented coverage
of inductance loop detectors, with a typical nominal spacing of
\unit[100]{m}.  In consequence, spatiotemporal patterns may
be identified without any interpolation process at all. Thus in
effect, the ground truth is directly available and we may use it to
definitively evaluate the performance of interpolation algorithms.

\begin{figure}[ht!]
\centering
\begin{tabular}{c}
\includegraphics[width=9cm]{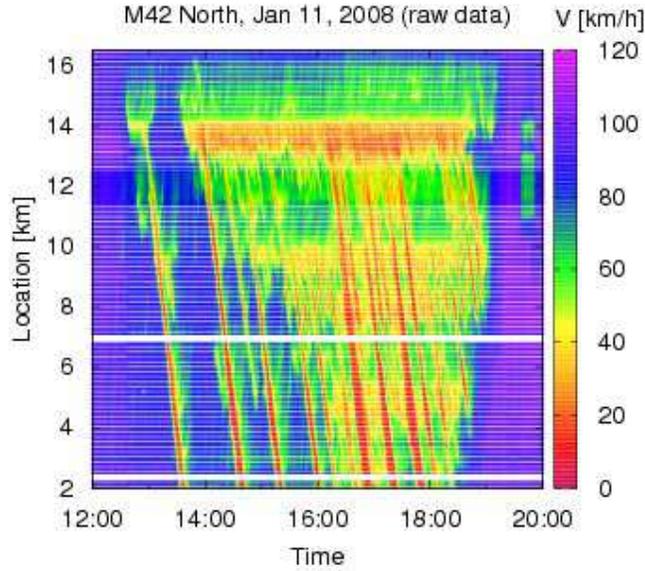} \\[-7mm]
\end{tabular}

 \caption{\label{fig:situation_M42} Reference situation used for
   validation. The inter-detector spacing is~\unit[100]{m}
   (\unit[40]{m} in the vicinity of $x=\unit[12]{km}$). The data are
   visualized as a spatiotemporal scatter plot. Each data point
   corresponds to the local speed aggregated over all lanes and
   over \unit[1]{min}. No further data processing has been applied. }

\end{figure}

As a test case, we take data from Friday January 11, 2008, and in
Fig.~\ref{fig:situation_M42} we display a scatter plot of 1-minute
aggregate lane-average speed, showing a complex spatially extended
pattern incorporating several bottlenecks and large amplitude
stop-and-go waves. Close examination of this raw data largely supports
the use of just two distinct propagation velocities $c_\text{free}$
and $c_\text{cong}$, validating the overall GASM technique.

\begin{figure}[ht!]
\centering
\begin{tabular}{cc}

\includegraphics[width=6.5cm]{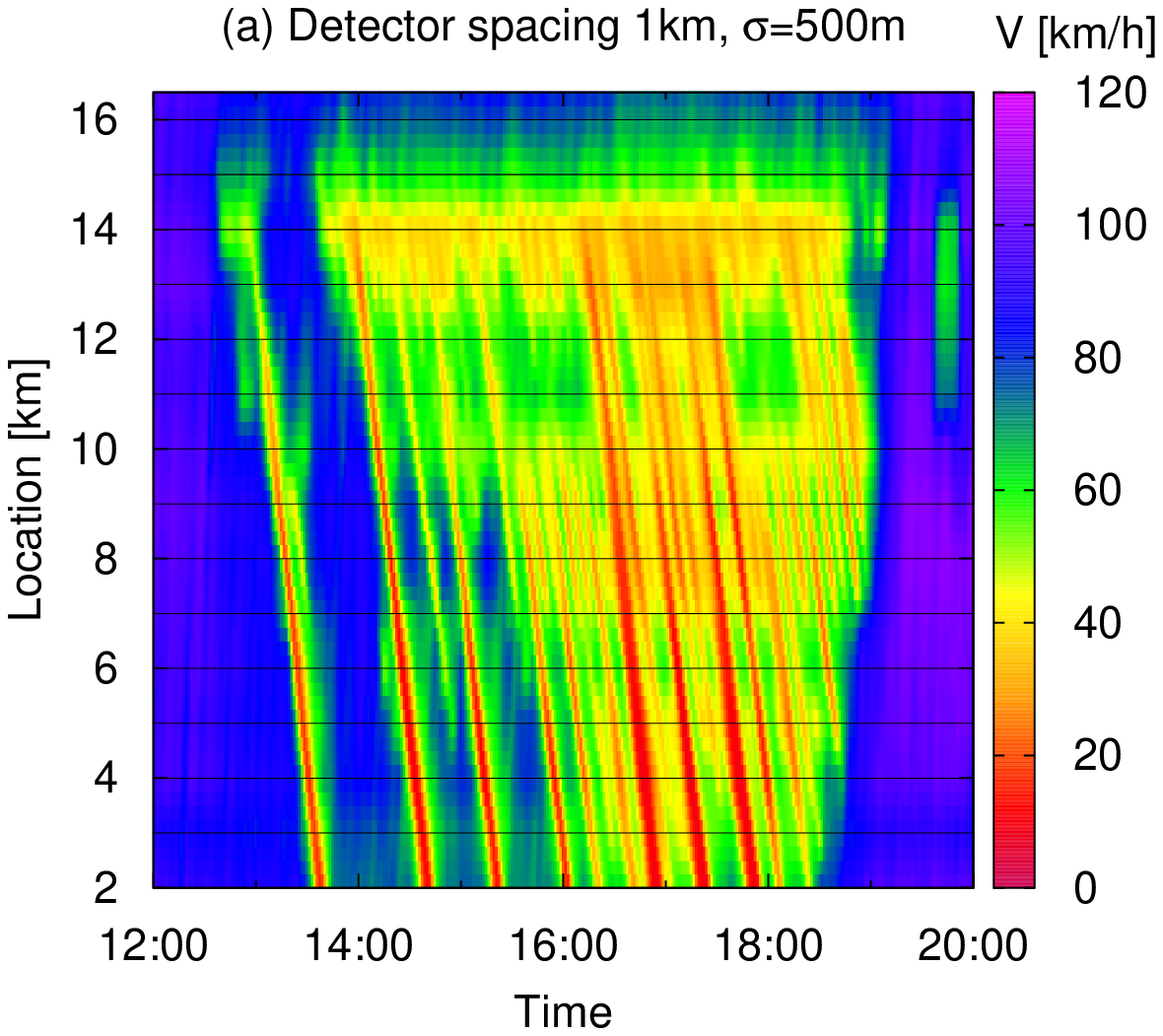} &
\includegraphics[width=6.5cm]{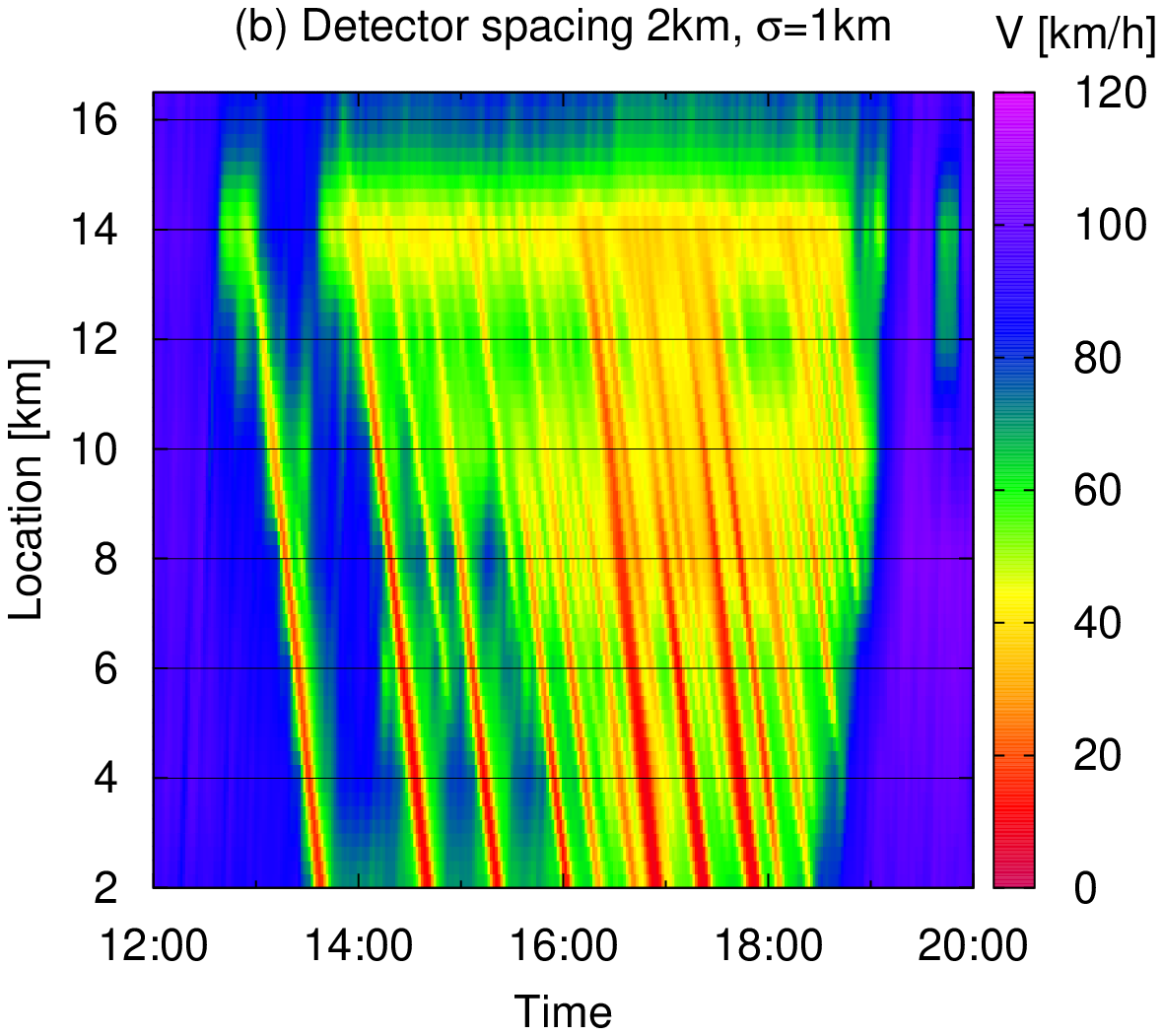} \\ [-4ex]
\includegraphics[width=6.5cm]{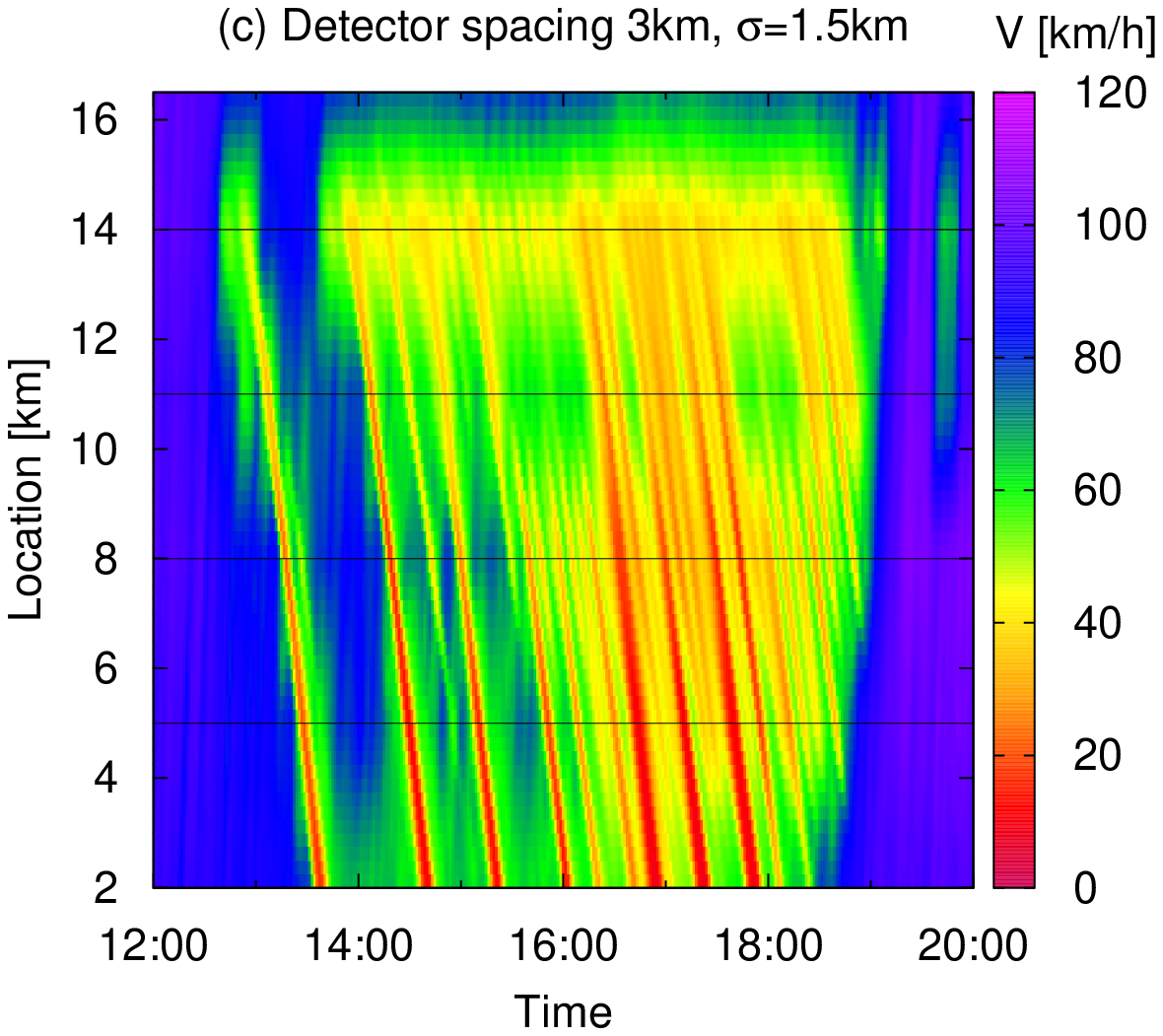} & 
\includegraphics[width=6.5cm]{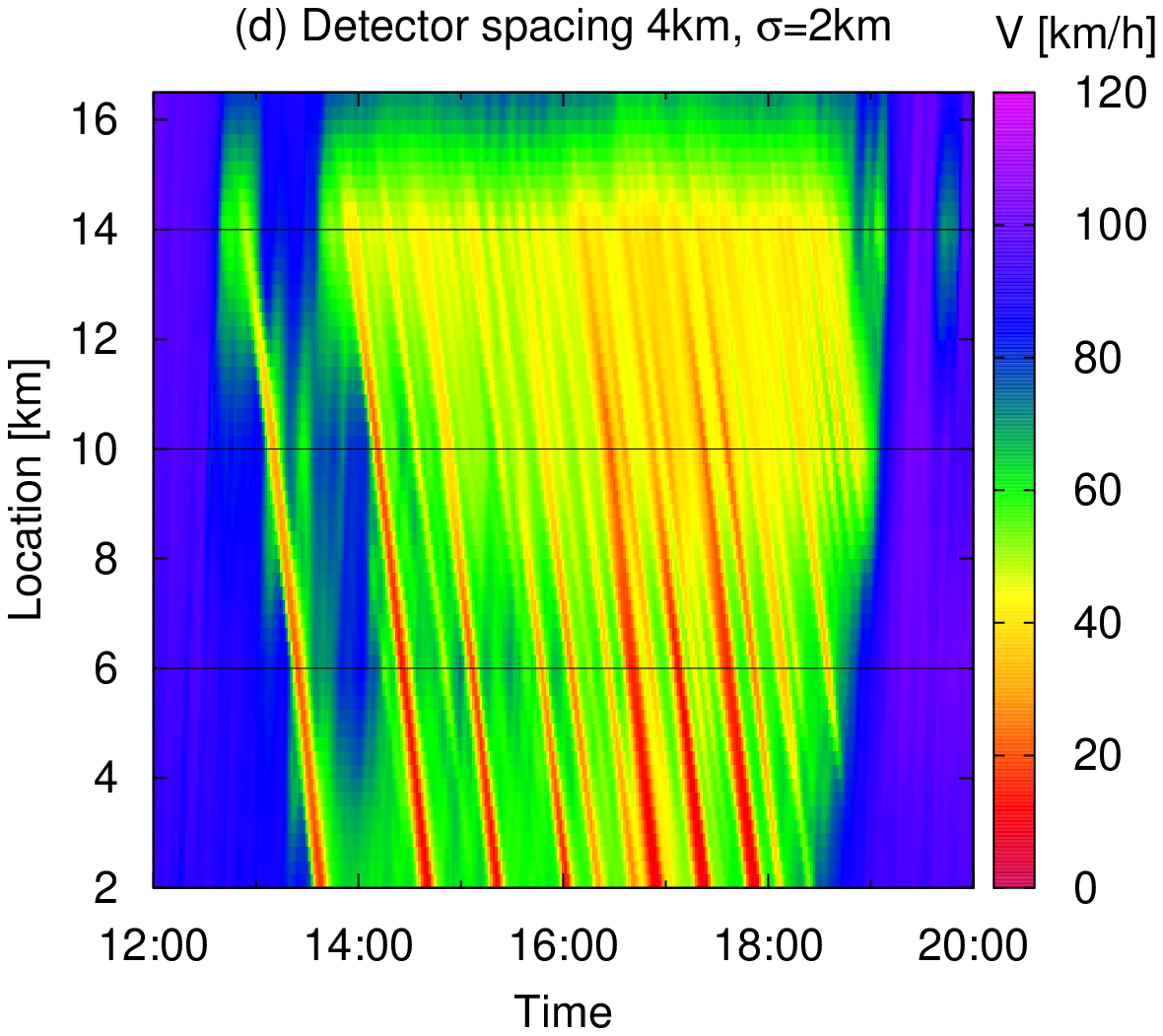} 
\end{tabular}
 \caption{\label{fig:validation_m42}Reconstruction of the reference
   situation of Fig.~\ref{fig:situation_M42} by the adaptive smoothing
   method (standard parameter set) applied on reduced data sets with
   detector spacings between \unit[1]{km} and \unit[4]{km}. The
   locations of detectors whose data has been used in the
   reconstruction are indicated by horizontal lines.}

\end{figure}

For a quantitative investigation, we apply the GASM with standard
parameters to input data chosen from just a small selection of the
available detectors. The interpolated field $V(x,t)$ is then compared
to speed data at detectors which are half way between those whose
data has been used in the reconstruction. For example, at a spacing of
\unit[1]{km} corresponding to Fig.~\ref{fig:validation_m42}(a), the
error measure is based on the detectors at $x=\unit[2.5]{km}$,
$x=\unit[3.5]{km}$ and so forth. Fig.~\ref{fig:validation_m42} displays
the reconstructed traffic states with reduced sets of loop
detectors. In summary, the most important features are identified even
when the detector spacing is increased to \unit[4]km.

\begin{figure}[ht!]
\centering

\includegraphics[width=6.5cm]{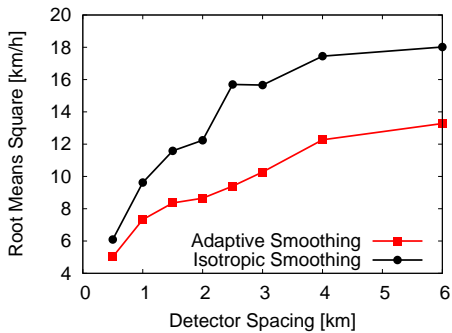}

 \caption{\label{fig:error_m42} Reconstruction of time series for
   speed using the adaptive smoothing method (lower curve) and
   conventional smoothing (upper curve). Shown is the RMS velocity
   deviation with respect to the actual measurement, averaged over all
   detector sites that are half way between the detectors available to
   the reconstruction methods. The deviation is plotted as a function
   of the spacing of the available detectors.}

\end{figure}

Fig.~\ref{fig:error_m42} presents RMS errors of the reconstructed
velocities, averaged over all applicable test sites, as a function of
the detector spacing. To assess the quality of the GASM, we compare it
with conventional isotropic smoothing, that is, setting
$c_\text{free}$ and $c_\text{cong}$ to $\infty$. For a given detector
spacing, the quality of the GASM can be compared to that of
conventional smoothing when about twice as many detectors are
available. Specifically, when using the GASM, the quality of the
reconstruction at a detector spacing of \unit[2.5]{km} is
comparable to that of isotropic smoothing when detectors are
available every kilometer.

\section{\label{sec:appl}APPLICATIONS}
We now give three illustrative examples of potential applications for
the GASM: 1.\ Compensation for detector failure
(Sec.~\ref{sec:dataGaps}); 2.\ Separation of noise from true traffic
dynamics (Sec.~\ref{sec:noise}); and 3.\ Fusion of floating car data (FCD)
with stationary detector data (SDD) (Sec.~\ref{sec:fcd}).

\subsection{\label{sec:dataGaps}Compensation for Detector Failure}
Common operational problems with inductance loop systems include
\begin{itemize}
\item a temporary but simultaneous failure of all detectors covering a
certain road section, usually due to a
failure in the communications sub-system; or
\item the permanent failure of one or a few detectors, often due to
  their installation not meeting specified standards.
\end{itemize}
In either case, the resulting SDD has `gaps', and
we may estimate the data that is missing by interpolating the data
that is extant. 

\begin{figure}
\centering
\includegraphics[width=\textwidth]{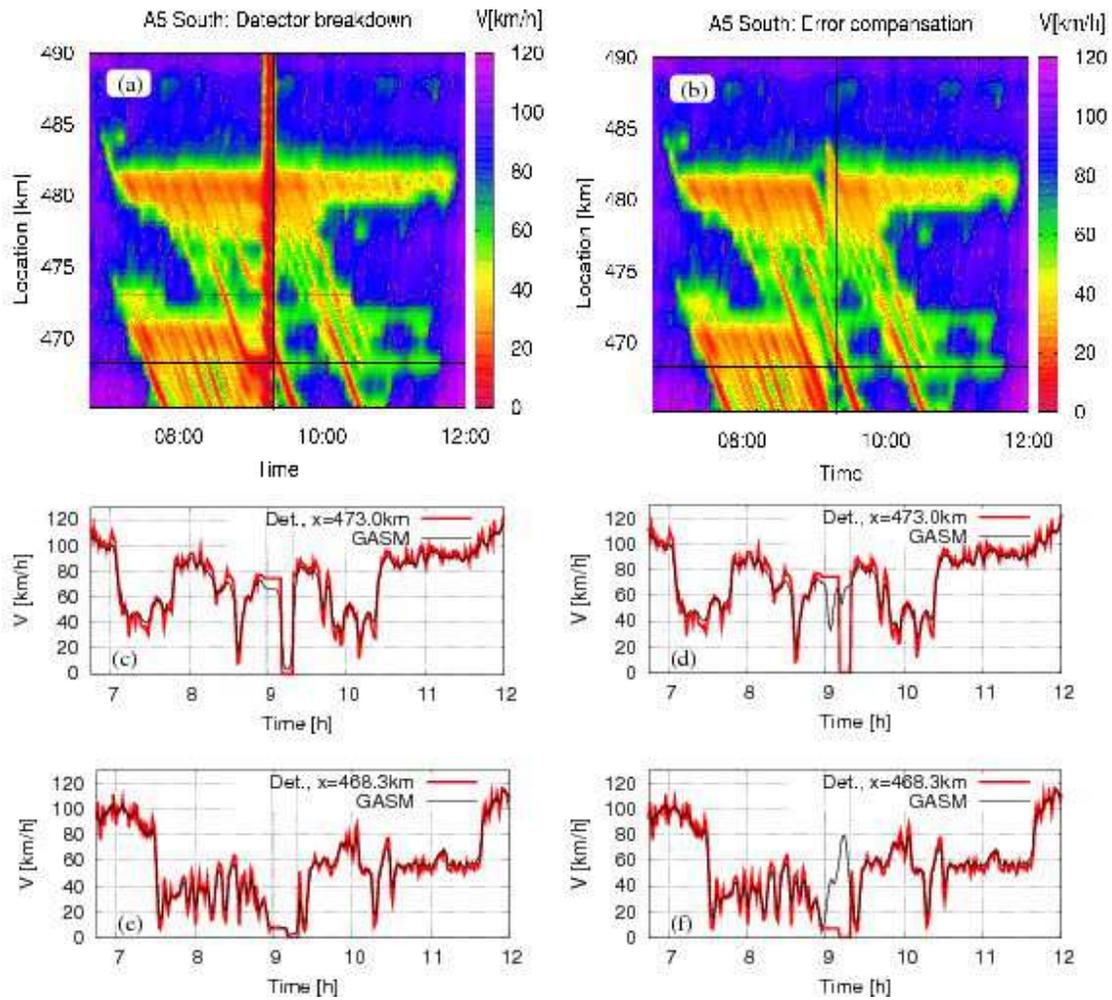} 
 \caption{\label{fig:det_error}Detector failure between 08:59 and
   09:19 (South-bound A5 autobahn on August 6th, 2001). Vertical lines
   in all plots indicate the period of failure. Plots (a,c,e) show
   results when GASM incorporates erroneous data and (b,d,f) shows the
   correct results when GASM omits the erroneous data and fills the
   gap by interpolation.}
\end{figure}

Our example is of the former type and is taken from the South-bound A5
autobahn near Frankfurt, Germany, on the morning of August 6th, 2001,
see Fig.~\ref{fig:det_error} and~\citep{Martin-empStates} for full
details of the site. There is a \unit[20]{min} breakdown of all
detectors between 08:59 and 09:19. During this period, the detection
system becomes `frozen' for 10 minutes and then records zero speed for
a further 10 minutes. Fig.~\ref{fig:det_error}(a) displays the
resulting spurious spatiotemporal field recovered by the GASM.

However, additional error bits signal the detector failure and the
correct approach is thus to eliminate the period 08:59 to 09:19 and
apply the GASM to what remains, see Fig.~\ref{fig:det_error}(b).
Observe that the GASM can bridge the gap in the data in a mostly
natural way, although an artefact is introduced in the transition from
congested to free traffic at the bottleneck at kilometer 482. This is
because the GASM is well-tuned to reconstruct missing information in
structures with velocities $c_{\rm free}$ and $c_{\rm cong}$, but not
to reconstruct stationary structures.

To further clarify the properties of the reconstructed state,
Figs.~\ref{fig:det_error}(c-f) display time series of speed at two
selected detectors. Each plot shows the original detector data, and
Figs.~\ref{fig:det_error}(c,e) compare it with the erroneous GASM
reconstruction whereas Figs.~\ref{fig:det_error}(d,f) compare with the
correct reconstruction with the erroneous data removed. Notice that 
the reconstructions in Figs.~\ref{fig:det_error}(d,f) do not
correspond to a simple interpolation in time, because of the way in
which the GASM incorporates spatial information from other detectors.

Further investigations have shown that temporal data gaps of up to
\unit[30]{min} and spatial gaps up to \unit[3]{km} can typically be
compensated for. In the latter case, however, the result depends
strongly on the position of the failed detectors relative to
bottlenecks.

\subsection{\label{sec:noise}Elimination of Noise}

A significant problem in SDD is noise due to the
time interval (typically 1 minute) over which raw data is aggregated
by the road-side signal processing subsystems. In particular, when the
traffic flow rate is low, the aggregate may be constructed from only
a handful of individual vehicle observations, which themselves are
prone to measurement error, and in consequence it is subject to gross
statistical sampling error. The two situations of concern are dense
queuing traffic and sparse high-speed traffic, and the latter case is
subject to the additional problem that the variance in individual
vehicles' true velocities may be very high.

A common real-time application is to use speed and flow thresholds
from single detectors to trigger queue warnings on Variable Message
Signs immediately up-stream of a stop-and-go wave. It is thus crucial
that stop-and-go waves are identified correctly and that alerts are
not triggered by statistical outliers.  To reduce the effect of
sampling error, one may aggregate data over wider time
intervals. However, there is not a clear separation of time scales and
sharp features such as the boundaries of the waves may get lost if the
aggregation window is broadened too far. Hence in general, it is
non-trivial to separate noise from dynamics in detector time series.

\begin{figure}
\centering

\includegraphics[width=80mm]{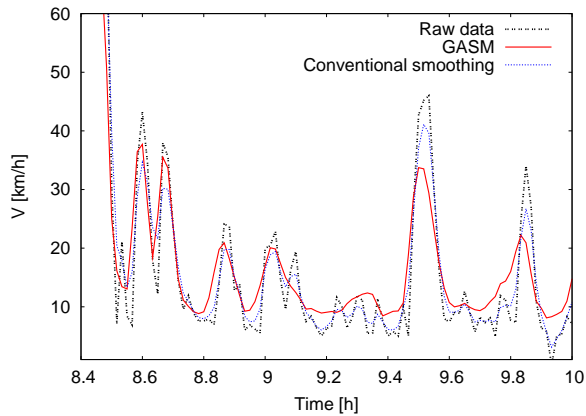}
\caption{\label{fig:noise}Time series for a stationary detector at
  location $x=\unit[473.0]{km}$ on the German Autobahn A5-South (see
  Fig.~\protect\ref{fig:det_error} for location --- but this data is
  taken from a different day).  The original data (thick solid line)
  is compared with the reconstruction at this position using the
  generalized adaptive smoothing method (thin solid), and with
  conventional smoothing (thin dashed line).}
\end{figure}

For illustration, we again consider the South-bound A5 autobahn near
Frankfurt, Germany \citep{Martin-empStates}, with data now taken from
July 9, 2001, on which there were a number of strong stop-and-go
waves. See Fig.~\ref{fig:noise}, which displays the time series
of speed from a single stationary detector and compares
conventional temporal smoothing using the kernel $\exp(-|t|/\tau)$
with $\tau=\unit[60]{s}$ and the GASM with standard parameters.

In summary, the GASM is much more effective than simple temporal
filters at reducing noise whilst retaining structure. In effect, the
GASM uses spatial information, by blending the time series of nearby
detectors, to enhance vehicle counts without broadening the temporal
averaging window. The difference between noise and real traffic
oscillations can thus be identified, because traffic information is
correlated between detectors whereas noise is not.

\subsection{\label{sec:fcd}Fusion of Floating Car Data}

As Intelligent Transport Systems are rolled out further across our
trunk road networks, savings in infrastructure costs may be achieved
if inductance loops are installed much more sparsely than is typical
for the busiest highways. However, a detailed picture of traffic flow
may still be derived if we also have access to some FCD, for example
from high-end GPS services (such as the ITIS~\citep{ITIS} or
Trafficmaster~\citep{trafficmaster} systems in the United Kingdom) or
from data from the mobile phone networks, that is so-called
floating-phone data (FPD) \citep{caceres2008review} which is
incorporated into high-end navigation services such as
TMCpro in Germany ~\citep{tmc-pro}.

As a final application of the GASM, we consider the fusion of FCD and
SDD. Our illustration uses synthetic data generated by microsimulation
based on the {\it Human Driver Model} (HDM) \citep{HDM}, which is a
refinement of the well-known {\it Intelligent Driver
  Model}~\citep{Opus}. In the HDM, the reaction-time and
multi-anticipatory effects have been calibrated so that the
quantitative details (wavelength, amplitude etc.) of spatiotemporal
traffic patterns are reproduced.

\begin{figure}
\centering
\includegraphics[width=80mm]{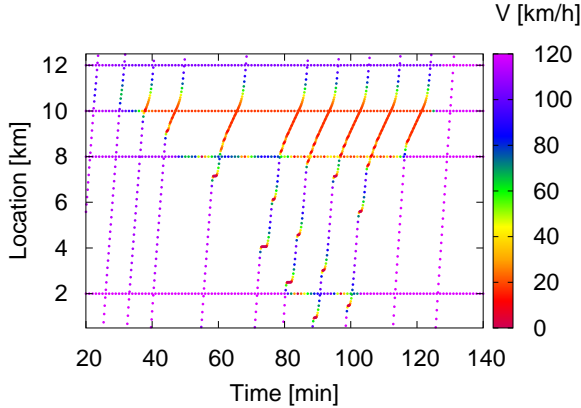}

\caption{Scatter plot of simulated Floating Car and Stationary
  Detector speed data.}
\label{fig:fusion1}
\end{figure}

\begin{figure}
\centering
\includegraphics[scale=0.6]{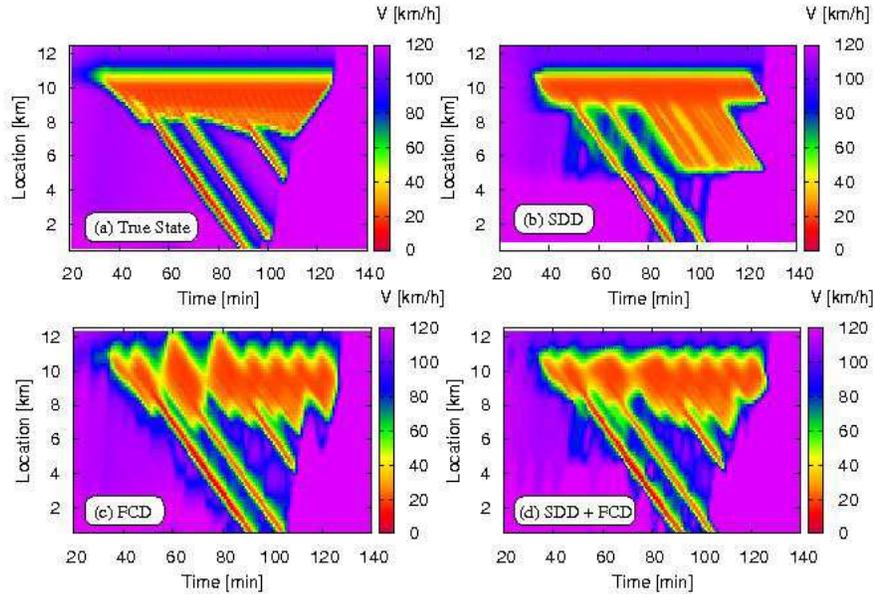}
\caption{Spatiotemporal speed profiles: (a) ground truth; (b) using
  Stationary Detector Data only; (c) using Floating Car Data only; (d)
combining Stationary Detector and Floating Car Data. }
\label{fig:caption2}
\end{figure}

Our specific example simulates a \unit[12]{km} section of a
single-lane highway without junctions over more than 2
hours. Rush-hour conditions are simulated by a peak in the in-flow
(traffic demand) at $x=\unit[0]{km}$. To induce traffic patterns, a
flow-conserving bottleneck is applied at $x=\unit[11]{km}$ by
increasing the car-following model's time-headway parameter at this
point. Four stationary detectors are then simulated at
$x=\unit[2,8,10,12]{km}$, and just 10 vehicles out of a total of
2,750 in the simulation are selected randomly to provide speed data
at \unit[10]{s} resolution. These SDD and FCD are presented in the
scatter plot in Fig.~\ref{fig:fusion1}.
We then reconstruct spatiotemporal velocity profiles by applying the
GASM to different combinations of the SDD and FCD data. See
Fig.~\ref{fig:caption2}. Here Fig.~\ref{fig:caption2}(a) displays the
ground truth profile derived by using FCD from all vehicles in the
simulation. Note that for the reconstruction based on SDD alone, not
only interpolation but also extrapolation is required at the extreme
upstream and downstream locations.

In Fig.~\ref{fig:caption2}(d) SDD and FCD are combined using the same
kernel function, except that each FCD point has been given double the
weight of an SDD point. In practice one would need to experiment with
this relative weighting and / or tune it to the particular application
under consideration. 

In summary, the SDD reconstruction outperforms FCD in resolving the
stationary downstream front at the bottleneck at $x=\unit[11]{km}$,
whereas the FCD reconstruction outperforms SDD in resolving
stop-and-go waves. The reconstruction using both FCD and SDD together
combines the best properties of either individual reconstruction.
This example shows that the incorporation of even very small amounts
of FCD (one vehicle in every 200 or 300) can significantly improve 
the reconstruction of the traffic state.

\section{\label{sec:disc}DISCUSSION}

In summary: in this paper we have demonstrated how the Generalised
Adaptive Smoothing Method (GASM) can solve practical problems in the
analysis of highway traffic patterns. In particular it can
1.\ compensate for gaps in data caused by detector failure; 2.\ reduce
noise due to either sampling or measurement error (whilst compromising
resolution less than a purely temporal filter); and 3.\ fuse
heterogeneous data. Although we have focussed on the reconstruction of
the spatiotemporal velocity field, other fields (e.g., flow) may also
be reconstructed: to achieve this, use alternative data in the linear
formulae \eqref{eq:asm_v_free} and \eqref{eq:asm_v_cong} but retain
unchanged the dependence on velocity in the nonlinear component
\eqref{eq:asm_super}, \eqref{eq:asm_w_filter}.

We have established that the performance of the GASM is quite robust
to changes in its parameters. In consequence, it does not need tuning
for each new scenario, and for freeway applications, the values from
Table~\ref{tab:asm_param} may be used with confidence. A substantial
benefit of the method is that it is not based on any one specific
model of highway traffic and therefore may be used in the objective
benchmarking of different models.  On the other hand, a
fine-tuning of the parameter $c_{\rm cong}$ may be used for research
purposes to determine the propagation velocity of congested traffic
patterns --- and when formulated in r.m.s.\ error this method is
related to the standard one based on the cross-correlation of time series 
from different stationary detector locations \citep{Zielke-intlComparison}.

However, the results here represent only a first analysis and a
detailed calibration and validation study remains for future work.  We
have yet to carry out a formal optimisation of the GASM parameters
(e.g., using over-specified inductance loop data, as from the English
M42 motorway), but there are other ways in which the details of the
formulation that we have presented may be varied, and they should be
analysed systematically using a variety of data from different
highways and different countries. For example, we might experiment
with different kernel functions (and note in practice these are
implemented with a finite range `cut-off' for computational
efficiency). Also, the relative weighting and parametrisation of the
kernels applied to different types of data point (individual vehicle
or aggregate) needs investigation, and if possible a rigorous
grounding in Bayesian statistics. 

In fact, in either wholly congested or wholly free traffic, the GASM
is linear and thus its performance may be compared definitively to a
more general space of linear filters tuned for any one highway
scenario. However, in practice, it is not so easy to generalise the
nonlinear switch \eqref{eq:asm_super}, \eqref{eq:asm_w_filter} which
is key to the reconstruction of complete traffic patterns.

Unlike the ASDA/FOTO method~\citep{Kerner-ASDA-FOTO}, the GASM in its
present form is not suited to online applications. In principle, our
method might be used to extrapolate data from the past into the near
future, and thus form the basis of a short-term forecasting
algorithm. In fact, stop-and-go waves are forecasted very well by this
approach. The problem is the behaviour of the GASM at bottlenecks
where there is commonly a congestion pattern whose downstream front is
stationary with respect to the highway \citep{Helb-Phases-EPJB-09}. At
present, the GASM does not distinguish different congested flow
regimes and so in extrapolation to the future, it will erroneously
propagate this structure away from the bottleneck at velocity $c_{\rm
  cong}$. The correction of this problem remains for future work, as
does the incorporation of other aspects of traffic physics which may
be useful in forecasting, such as the conservation of vehicles.

From the point of view of infrastructure design and specification, it
is interesting to investigate the optimal positioning of a fixed
number of stationary detectors, in order for the GASM to yield the
best reconstruction of the velocity field. To answer this question we
should again return to the performance of the method in an environment
like the M42 motorway where there is a redundancy of detectors. We
suspect that detectors should be placed more densely near bottlenecks,
but the optimal configuration is unknown.

Finally, the increasing availability of Floating Car Data (FCD) may
allow much coarser stationary detector data (SDD) in future. A
systematic investigation of the effects of average SDD spacing, SDD
aggregation time, FCD penetration rate, and FCD aggregation time /
event-based protocols remains for future work.

\paragraph{Acknowledgments}
The authors thank the Verkehrszentrale Hessen and the
Autobahndirektion S\"udbayern for providing German autobahn data and
the English Highways Agency for access to MIDAS motorway data sets.
R.E.~Wilson acknowledges the support of EPSRC Advanced Research
Fellowship EP/E055567/1.


\end{document}